\documentclass{article}
\usepackage{graphicx}
\def\ben{\begin{equation}}
\def\een{\end{equation}}

\def\bea{\begin{eqnarray}}
\def\eea{\end{eqnarray}}

\newcommand{\labeq}[1] {\label{eq:#1}}

\input amssym.def
\input amssym.tex

\begin{document}

\hfuzz=100pt
\title{The Measure Problem in Cosmology}
\author{G.~W.~Gibbons and Neil Turok
\\
DAMTP,
\\ Cambridge University,
\\ Wilberforce Road,
\\ Cambridge CB3 0WA,
 U.K.
}
\maketitle
                                                   
\begin{abstract} 
The Hamiltonian structure of general relativity provides a natural canonical measure on the space of all classical universes, {\it i.e.}, the multiverse. We review this construction and show how one can visualize the measure in terms of a ``magnetic flux" of solutions through phase space. Previous studies identified a divergence in the measure, which we observe to be due to the dilatation invariance of flat FRW universes. We show that the divergence is removed if we identify universes which are so flat they cannot be observationally distinguished. The resulting measure is independent of time and of the choice of coordinates on the space of fields. We further show that, for some quantities of interest, the measure is very insensitive to the details of how the identification is made. One such quantity is the probability of inflation in  simple scalar field models. We find that, according to our implementation of the canonical measure, the probability for $N$ e-folds of inflation in single-field, slow-roll models is suppressed by of order ${\rm exp} (-3N)$ and we discuss the implications of this result.
\end{abstract}

\section{Introduction}

The problem of comparing different possible histories of the universe, and assigning a probability to each, is central to theoretical cosmology.  We cannot expect a fundamental theory to predict precisely what we see today: at best, it should predict an ensemble or ``multiverse" of possible universes, with the universe we observe being a typical member.

Of course there are many attitudes that one may take to probability
not only  in the context of cosmology, but in science more generally. 
The stance adopted in this paper  is that in practising science we
often adopt Bayesian methods which make essential use of a priori
probabilities (even if one does not accept this, or has doubts about what
these probabilities mean, this is certainly a fair characterisation
of much of the current observational literature).

In fact, following Laplace's Principle of Indifference~\cite{Laplace}, we often start with the least informative probability distribution, {\it i.e.}, a flat distribution, and then sharpen it in the light of additional information. This is,  as we see it, the essence of the Bayesian approach. It requires a well defined a priori measure, {\it i.e.}, one with finite total measure.

At the risk of falling into pedantry let us spell this out more formally. We define 
the {\it a priori probability} $P(U)$ to be the probability that the universe is $U$, according to some fundamental theory. Likewise we define $P(O)$ to be the probability of making an observation $O$ in {\it any} universe. The joint probability that the universe is $U$ and that we make an observation $O$ is $P(U\cap O)$. The conditional probability of making an observation $O$ in a universe $U$ is $P(O|U)$ (called the {\it likelihood}), and the conditional probability that we are in the universe $U$ given that we have made an observation $O$ is $P(U|O)$ (called the {\it a posteriori probability}). It follows from elementary considerations that
\ben
P(U|O) P(O)= P(U \cap O) = P(O|U)P(U)
\een
whence the  Cosmic Bayes's Theorem~\cite{Bayes}  
tells us that 
\ben
P(U|O)={P(O|U)P(U) \over \int_M P(O|U)P(U) dU }
\label{aa1} 
\een
where the integral is over what we call the {\sl multiverse}, $M$, and 
$dU$ is a measure on the multiverse. Equation (\ref{aa1}) describes how the a priori probability $P(U)$ is updated in the light of observations.

In this paper, we define the multiverse $M$ to be the set of all possible universes, or, better, the set of all model universes. These are by definition disconnected from one another. {\sl  This differs from other popular interpretations} in which the multiverse is a connected spacetime containing many causally disconnected regions and, roughly speaking, probabilities are taken to be proportional to the numbers of such regions. This latter concept has been termed the {\it metauniverse}~\cite{Moss,Vilenkin}. It is not obvious to us that this second meaning can be made quantitively precise in such a way that the probabilities are well defined. The basic problem is that the natural measure on spacetime, $\int d^4 x \sqrt{-g}$, is usually infinite in the scenarios being considered, and the infinity is not removed by restricting attention to so-called reheating surfaces. Much of the current literature on inflation appears to reflect this difficulty. 
   
By contrast to Laplaces's  Principle, one may regard a Proposal for the State of The Universe as some choice of probability distribution function $P(U)$ relative to the a priori measure $dU$. It might possibly constitute an ``explanation" if derived from some underlying fundamental theory which, again, must be well defined to be meaningful. {\sl Note that in this paper we are not setting out to make a specific proposal for the universe, i.e.,} we are not advocating a specific formula for $P(U)$, but rather, as we shall argue in detail later, we are adopting
Laplace's Principle of Indifference and finding that indeed this is inadequate to 
favour inflation. A more specific, sharper probability distribution is required to explain inflation. One may quantify the sharpness of a probability distribution by defining (following von Neumann, Shannon and others) the {\it information}
\ben
I=\int_M P(U) \ln \left( P(U)\right) dU,  
\label{aa2}
\een
which, after imposing the constraint $\int P(U) dU =1$, is minimised by the uniform distribution $P(U)=$constant, as one can easily check by computing the first and second variational derivatives and using $P(U)\geq 0$. The quantity $I$ measures the information (or lack of randomness) carried in a typical realisation of $P(U)$: the least information is obtained if all possible outcomes are equally likely, and the greatest if a single outcome is certain. The {\it information entropy} $S$ is then defined to be the negative of $I$, which is of course maximised in the uniform distribution. As is well-known, maximising $S$ under various additional constraints including the conservation of energy yields the usual formulae of statistical mechanics. 

In this paper, we show that, with the canonical definition of $dU$ given by Hamiltonian dynamics, Laplace's principle does not typically predict inflation. Hence, even if a fundamental theory allows inflation, a sharper $P(U)$is required in order to explain why inflation actually occurred. It remains as a challenge to fundamental theory to explain how such a $P(U)$ might arise. The predictiveness of any particular proposal can be assessed from (\ref{aa2}); what we would be most happy with is a highly predictive theory, whose predictions were consistent with observation. In fact this idea can be taken further, as we have been  reminded by Don Page. If we regard a proposal $P(U)$ as a {\it hypothesis}  then we may, given an appropriate measure  on the infinite dimensional set of hypotheses  $\{P(U)\}$, adopt Bayesian methods to evaluate their relative probabilities.  We shall not attempt this formidable task in this paper. 

The ideas sketched above are not new, and many people have pursued
them before in some, largely qualititive, way. In this paper we attempt 
, following \cite{GibbonsStewart}, to make them quantitative.  In fact our own viewpoint was influenced in part by trying to make some of Penrose's arguments \cite{Penrose1,Penrose2,Penrose3} mathematically precise.

There are several additional problems peculiar to cosmology when attempting to construct a statistical theory. First, there there is the problem of general covariance. There is no absolute notion of space or time in general relativity: these are properties of each particular classical solution of the field equations. Second, the solutions generically possess singularities in the past or the future, where the field equations break down in finite time.  Even if one makes the drastic simplification of restricting attention to Friedmann-Robertson-Walker (FRW) universes, {\it i.e.}, so-called mini-superspace, it is not obvious how to compare different classical spacetimes which are solutions to the same laws of physics. Finally, in theories like string or M-theory, where the low energy effective description involves many additional fields, there may not even be a preferred space-time metric: one can change coordinates on the space of fields (for example via Weyl transformations) in an arbitrary way and physical results should not depend on these choices.

In spite of these difficulties, the problem of constructing sensible measures on the space of solutions is of undeniable importance to the evaluation of various cosmological scenarios. Most of these scenarios focus on some subset of classical solutions, for example arguing that they are ``generic" because they exhibit dynamical attractor behavior. But to understand how predictive the proposals really are, in the absence of any further information, we need to quantify the extent to which the flat probability measure, without any additional input, succeeds in narrowing down the range of allowed classical solutions. 

In recent times, the problem of the measure in cosmology has grown hugely in importance. The claimed vast ``landscape" of possible compactifications and moduli in string theory seems to offer a truly bewildering range of classical cosmologies\cite{DouglasKachru}. When combined with recent observations of dark energy, the landscape picture has encouraged many physicists to pursue anthropic explanations. However, the lack of an {\it a priori} measure on the space of model universes is a serious flaw in such attempts, as even the most ardent proponents will admit. Nevertheless, remarkably little attention has been devoted to constructing measures whereby the different possible classical spacetimes within the ``multiverse" can be compared \cite{Coule,Kirchner}. In a similar vein to the anthropic arguments, a "top down" approach to cosmology has been proposed by Hawking and Hertog~\cite{HertogHawking}, in which one imposes constraints on the state of the universe today and attempts to determine its most probable history, for example by constraining the initial state to satisfy Hartle and Hawking's no-boundary proposal~\cite{HartleHawking}. However, it is hard to judge the success or otherwise of this approach without being able to quantify the extent to which it narrows down the multiverse.

Nearly two decades ago, Gibbons, Hawking and Stewart~\cite{GibbonsStewart} identified a natural measure on the set of classical cosmological solutions. They argued that any such measure should satisfy the following requirements: (i) it should be positive, (ii) it should depend only on the intrinsic dynamics and neither on any choice of time slicing nor on the choice of dependent variables, and (iii) it should respect all the symmetries of the space of solutions without introducing any additional ad hoc structures (e.g. ``Planck mass cutoffs") not arising from the field equations themselves. They showed that a measure satisfying all these requirements arises naturally from the Hamiltonian structure of general relativity and that it can be used to count the number of different solutions of the classical field equations in a well-defined way. Note that the set of all Cauchy data is unsuitable for this purpose because different Cauchy data can yield the same classical solution, evaluated on a different time-slice. The proposed measure avoids this overcounting problem by counting each distinct classical solution, in its entirety, only once.

One obvious application of such a measure on the multiverse is in determining the probability of inflation, {\it i.e.}, how likely is it that the universe started out in an inflationary state, within different models of the laws of fundamental physics. This question is potentially an Achilles' heel for inflation: if inflation is itself highly improbable, then it cannot be claimed to solve the classic cosmological fine tuning puzzles, of large-scale homogeneity, isotropy and flatness. Initially, it seemed that the canonical measure did indeed favor inflation~\cite{GibbonsStewart}, in simple inflationary models, but more detailed investigations, in particular by Hawking and Page~\cite{HawkingPage}, identified a serious ambiguity. They found that the canonical measure is infinite, and that both inflationary and non-inflationary solutions typically have infinite measure. Unfortunately, this left the problem of whether inflation is likely, or unlikely, unresolved. Since then, the problem has not received much attention, although various intuitive arguments, generally not satisfying the three above-mentioned conditions, have been given for or against inflation.

In this paper, we shall revisit the canonical measure, showing that when treated with sufficient care, a finite measure, still satisfying the Gibbons-Hawking-Stewart conditions, can be obtained. In agreement with Hawking and Page, we find the canonical measure is infinite. However, we identify the divergence as being due to the dilatation symmetry of flat FRW universes, and we show it is removed if one identifies solutions which cannot be observationally distinguished. Concretely, we impose a limit on the parameter known in the observational literature as $\Omega_k$, measuring the ratio of the space curvature term in the Friedmann equation to the square of the Hubble parameter. If all universes with $|\Omega_k|$ smaller than some observational limit $\Delta \Omega_k \ll 1$ are identified, then the canonical measure becomes well-defined. Furthermore, provided the limit on $|\Omega_k|$ is imposed once the matter fields have entered a phase of evolution in which the expansion of the universe acts adiabatically, {\it i.e.}, as a slow variation of parameters in the matter Lagrangian, our measure reduces to the canonical measure on the matter fields alone, and is an adiabatic invariant. 
In this regime the measure becomes independent of the conjugate ``angle" variable. This statement is independent of the time, the values of the cosmological parameters, or indeed the value of $\Delta \Omega_k$ when the cutoff is imposed. These good properties motivate us to reconsider the probability of inflation with various numbers of e-foldings, using the proposed canonical measure. The probability of $N$ e-folds of inflation depends critically on the ``angle" variable, and we find that the probability of obtaining $N$ e-folds of inflation in simple inflationary models is suppressed by a factor of ${\rm exp} (-3 N)$. Since $N > 50-60 $ is typically required in realistic models (depending on the efficiency of reheating), it follows that inflationary solutions are, in fact, tremendously rare in the space of classical solutions. We emphasize these statements are independent of the details of the cutoff imposed on the canonical measure.

Our conclusion is, of course, at variance with much of the
inflationary literature over the last two decades.  It has some
resonance with the well-known arguments (see
e.g. Refs.~\cite{Stewart,Penrose1,Penrose2,Penrose3}) that inflation cannot possibly solve the classical cosmological puzzles for arbitrary initial conditions because {\it any} physically permissible current state of the universe would, if run back in time, correspond to {\it some} initial conditions, and also with Holland and Wald's arguments~\cite{HollandsWald} that any canonical measure would naturally assign the same probability to deflation as to inflation, since Einstein's equations are time-reversal invariant. Because the canonical measure is a measure on the set of universes, independent of their time-orientation, we do indeed find this result: both inflation and deflation are predicted to be exponentially rare phenomena among the set of all classical solutions of the field equations.

The structure of this paper is as follows. In Section 2 we review the construction of the canonical measure, clarifying the geometrical structure in elementary terms and in particular demonstrating that the measure is independent of the choice of initial slice in the space of dynamical variables (contrary to the apparent assertion of Hollands and Wald on this matter~\cite{HollandsWald}). In Section 3, we discuss the classical dynamics of simple scalar field models and we compute the canonical measure for scalar field matter in FRW spacetimes with arbitrary spatial curvature. At face value, the measure exhibits a divergence at large scale factor, {\it i.e.}, in the flat space limit, but we argue that since the scale factor is unobservable in this limit, one must factor out dilatations from the result. We show that this results in a sensible measure on the space of classical solutions, {\it i.e.}, one of finite total measure. In Section 4 we solve the classical field equations, identifying the solutions undergoing $N$ e-folds of inflation and computing the associated measure.  Section 5 compares our results with those obtained in previous discussions, and explains why our conclusions are so different from those reached on the basis of intuitive reasoning from ``chaos at the Planck density". Section 6 concludes.

\section{Liouville measures and magnetic flux} 

The principles behind the canonical measure are extremely simple. If one restricts attention to consistent (in the technical sense) finite-dimensional truncations, so-called mini-superspace models, then we are faced with a standard problem in Hamiltonian mechanics, and it is natural to bring to bear on the problem the standard techniques of statistical mechanics which have been used so successfully in all other areas of physics. 

The Einstein-matter equations provide a Hamiltonian flow in a  $2n$-dimensional  phase space $P$, equipped with a symplectic form $\omega$ which may be written in local Darboux coordinates as
\ben
\omega = \sum_{i=1}^n dp_i \wedge dq^i\,.
\label{eone}
\een  
The $n$'th power of $\omega$ gives the Liouville volume element on $P$
\ben
{ (-1)^{n(n-1)/2} \over n!} \omega ^n = d^n p \, d^n q\,.
\label{etwo}
\een
However, this is not what we want. We want to describe the set of distinct dynamical trajectories or, equivalently, the set of classical initial conditions giving distinct histories. In general relativity, the Hamiltonian ${\cal H}$ is constrained to vanish so the trajectories all lie on a $2n-1$ dimensional constraint submanifold
\ben
C = {\cal H}^{-1}(0)\,.
\label{ethree}
\een   
The set of classical trajectories is the Marsden-Weinstein quotient, also sometimes called the reduced phase space. This is the multiverse,
\ben
M=C/{\Bbb R} ={\cal H}^{-1}(0)/{\Bbb R},
\label{efour}
\een 
where ${\Bbb R}$ is the Hamiltonian flow. As Gibbons, Hawking and Stewart showed~\cite{GibbonsStewart} (see also \cite{Henneaux}), a symplectic form on the multiverse $M$ is naturally inherited from $\omega$: one simply chooses local coordinates in which $p_n={\cal H}$, from which Hamilton's equations imply that $q^n=t$, the time. As a result, one has 
\ben
\omega = \sum_{i=1}^{n-1} dp_i \wedge dq^i + d{\cal H}\wedge dt ,.
\label{efive}
\een  
so that the restriction to the constraint surfaces, ${\cal H}=0$, naturally yields a two-form on the space transverse to the Hamiltonian flow, 
$\omega_{C} \equiv \omega|_{{\cal H}=0}$. We now construct a measure on $M$ by raising $\omega_{C}$ to the $(n-1)$'th power:
\ben
\Omega_M \equiv { (-1)^{(n-1)(n-2)/2} \over (n-1)!} \omega_{C}^{n-1}.
\label{esix}
\een
Gibbons, Hawking and Stewart showed that the flux obtained by integrating this form is positive if measured in the correct sense, independent of slicing in phase space and invariant under Hamiltonian flow. We shall review this argument in more elementary terms below. Furthermore, (\ref{esix}) is natural in the sense that it requires no new elements in the theory other than those already present in the classical equations of motion. By its construction, it is invariant under any additional canonical symmetries. Hence, conditions (i)-(iii) given in the introduction are satisfied. 
%For an illuminating discussion of the problems encountered
%when using a non-canonical measure, see \cite{HollandsWald}.

One can visualize the measure (\ref{esix}) as the ``flux" of a divergence-free ``magnetic field". In general coordinates on phase space, the symplectic form $\omega$ is a covariant second rank anti-symmetric tensor field with components satisfying 
\ben
\omega _{\mu \nu}=-\omega_{\nu \mu}, 
\label{eseven}
\een 
where $\mu,\nu = 1\dots 2n$, and ${\rm det}\omega\neq 0$.
The symplectic form is closed,
\ben
d \omega =0,
\label{e8}
\een
which implies 
\ben
\partial _{[\mu} \omega _{\nu \tau]} =0,
\label{e9}
\een
and Hamilton's equations are 
\ben
V^\mu = \omega ^{\mu \nu} \partial_\nu {\cal H},
\een
with the summation convention, where $\omega ^{\mu \nu}$ is the inverse of 
$\omega _{\mu \nu}$ and the velocity on phase space, $V^\mu \equiv (dx^\mu / dt)$. 
Alternately, we may rewrite Hamilton's equations as
\ben
\omega _{\mu \nu } V^\nu = \partial _\mu {\cal H},
\label{e10}
\een
from which one obtains
\ben
V^\mu  \partial _\mu {\cal H}  = 0,
\label{e11}
\een
so the flow $V^\mu $ lies in the surfaces ${\cal H} = {\rm constant}$.

Now let us  choose coordinates such that
\ben
  x^{2n}  ={\cal H}.
\label{e12}
\een
The closure condition, restricted to ``spatial" indices, corresponding to directions tangent to the constraint manifold, is
\ben
\partial _{[a} \omega _{b c]} =0,
\label{e13}
\een
where $a,b =1,\dots,2n-1$, and we have
\ben
V^{2n} =0,
\label{e14}
\een
since the Hamiltonian has no explicit time-dependence. Thus from (\ref{e10}), and the fact that all ``spatial" derivatives of ${\cal H}$ are zero, we have $V^a \omega _{ab}=0.$

It is simplest to see what this means in the example $n=2$,  for which
$a,b,c$ run from 1 to 3. Define a magnetic field by
\ben
B_a \equiv {1\over 2} \epsilon_{abc} \omega _{bc},
\label{e15}
\een
then because $\omega$ is closed, ${\bf B}$ is divergence-free,
\ben
\partial_a B_a=0.
\label{e16}
\een
Moreover,
\ben
\epsilon_{abc} B_b V_c =0,
\label{e17}
\een
thus ${\bf V} $ is parallel to ${\bf B}$. In these formulae, we have lowered indices using the Kronecker delta, $\delta_{ab}$. Now, elementary arguments using the divergence theorem establish the point that the flux through some fixed surface is unchanged if the surface is deformed while keeping its boundary fixed, andunchanged if the surface is propagated forwards with the flow. Note that we have introduced $\delta_{ab}$ and $\epsilon_{abc}$, and are only dualizing only for convenience: we are merely implementing the standard rules of exterior calculus in a particular coordinate system and the key results, that the flux is conserved and that the flow lies parallel to the magnetic field are independent of the coordinate system and do not require a metric. 

The argument easily generalizes to higher dimensions, with the magnetic field constructed by dualizing the $(n-1)$'th power of the Hamiltonian symplectic form. 
Let 
\ben
B_a =  
      {1\over 2^{n-1} (n-1)!} \epsilon_{abcde\dots gh} \omega_ {bc} \omega _{de} \dots \omega _{gh}. 
\een
Then
\ben
\partial_a B_a =0
\een
follows from the closure of $\omega$ and
\ben
\omega_{[bc} \omega_{de} \dots \omega_{gh]} =
\epsilon_{abcde\dots gh} B_a
\een
and so
\ben
\epsilon_{abcde \dots gh} B_a V_b=0,
\een
which implies that
\ben
B_{[a} V_{b]} =0,
\een
that is, $V_a$ is parallel to $B_a$.

We end this section with a final point. Since $\omega$ is closed, it follows $\omega_C$ is also closed on $C$ and that the measure on the multiverse $\Omega_M\propto \omega_C^{n-1}$ is also closed. Hence one can always locally write $\Omega_M = d A$, with $A$ some $n-2$-form, arbitrary up to ``gauge transformations" $A\rightarrow A+d \chi$. However, the canonical construction gives more than this: $\omega$ may be written as $d(p_i dq^i)$ globally. So there is a natural definition of the ``vector potential" $A$, allowing one to reduce the $n-2$-dimensional integral of $\Omega_M$ over some surface $S$ to an $n-3$-dimensional integral of $A$ over the boundary of $S$. The latter integral may then be regarded as giving the integrated probability measure for all trajectories passing through $S$ or any topologically equivalent surface with the same boundary.

\section{Gravity and a Scalar Field}

In this section we consider a single minimally-coupled scalar field $\phi$ with potential $V(\phi)$ in a homogeneous and isotropic (FRW) universe. Generalizing the discussion to to additional scalar fields and other fields and fluids, or anisotropic cosmologies should be straightforward~\cite{Page1,Page2}. But the simplest case is interesting enough that we shall devote the remainder of this paper to it. A version of this model with two scalar fields, and its statistical properties, was analyzed by Starobinsky from a different point of view~\cite{Starobinski}. 
With the above-mentioned symmetry restrictions, the line element is
\ben
-N^2 dt^2+ a^2(t)\gamma_{ij} dx^i dx^j,
\label{b1}
\een
where $\gamma_{ij}$ is a metric on a space of constant (three-dimensional) scalar curvature $k= 0$ or $\pm 1$. Choosing units in which $M_{Pl}^2 \equiv 1/(8 \pi G) =1$, the Einstein-scalar action is 
\ben
{\cal S} = \int dt N\left( -3 a (N^{-2}{a'}^2-k) +{1\over 2} a^3 N^{-2} {\phi'}^2 -a^3 V(\phi) \right), 
\label{b2}
\een
where primes denote $t$ derivatives. 

Varying the action with respect to the lapse function $N$ yields the usual Friedmann equation
\ben
H^2 = {1\over 3} \left({1\over 2} \dot{\phi }^2 +V(\phi)\right) - {k\over a^2},
\label{b3}
\een
where dots denote proper time derivatives, with $d \tau = N dt$, and $H=\dot{a}/a$ is the expansion rate or Hubble parameter. Varying with respect to $\phi$ yields the scalar field equation
\ben
\ddot{\phi}+ 3 H \dot{\phi} = - V_{,\phi}.
\label{b4}
\een
Taking the time derivative of (\ref{b3}) and using (\ref{b4}) then yields
\ben
\dot{H}= -{1\over 2} \dot{\phi}^2 + {k\over a^2}.
\label{b5}
\een
Finally, varying with respect to $a$ yields a linear combination of 
(\ref{b3}) and (\ref{b5}). Equation (\ref{b5}) will be of particular interest to our later discussion. For $k\leq 0$, the Hubble parameter $H$ never increases, so no classical trajectory can cross a $H$=constant hypersurface more than once. Likewise, from (\ref{b3}), if $V$ is non-negative, then for $k\leq 0$, $H$ can never change sign. This implies that no classical trajectory can cross an $a=$ constant surface more than once.  

We now turn to the Hamiltonian analysis. The canonical momenta conjugate to $a$, $\phi$ and $N$ are
\ben
p_a= -6 a \dot{a}= -6 a^2 H, \qquad p_\phi = a^3 \dot{\phi}, \qquad p_N=0,   
\label{b5a}
\een
and the Hamiltonian is 
\ben
{\cal H} = N \left(-{p_a^2 \over 12 a }+{1\over 2} {p_\phi^2\over a^3} + a^3 V(\phi) - 3 ak \right),
\label{b5b}
\een
which vanishes by the equation of motion for $p_N$. We can use this to eliminate one of the four canonical variables $a,p_a,\phi,p_\phi$. Since the Hamiltonian is quadratic in both $p_a$ and $p_\phi$, either may be easily eliminated. Both choices have some merit, as we shall discuss below.

The first choice, used in previous treatments, is to eliminate $p_a$. This has the advantage that it is easily generalized to many matter fields, whether or not they have minimal kinetic terms. It proceeds as follows. From the vanishing of (\ref{b5b}) we
obtain 
\ben
p_a=\pm \sqrt{6 p_\phi^2 a^{-2} +12 a^4 V(\phi)-36 a^2 k}.
\label{b5ca}
\een
The choice of sign will occur in many subsequent formulae, but leads to no ambiguity. All it means is that at generic values of the remaining variables $a,\phi$ and $p_\phi$, there are two possible solutions, representing an expanding or a contracting universe. 

It is now convenient to change non-canonical coordinates to $\phi, \dot{\phi}$ and $\lambda\equiv \ln a$. The velocity of the flow in these coordinates is easily obtained:
\ben
V^a=\left(\dot{\phi}, -3 H \dot{\phi} -V_{,\phi}, H\right).
\label{b5da}
\een
The canonical two-form on the constraint manifold $C$,
\ben
\omega_{C}=\left(dp_a\wedge da +d p_\phi \wedge d\phi\right)|_{{\cal H}=0}.
\label{b6a}
\een
is straightforwardly evaluated in the same coordinates,
\ben
\omega_{C}=
e^{3 \lambda}\left(-(\dot{\phi}/H)d \dot{\phi}\wedge d\lambda + (3 \dot{\phi}+V_{,\phi}/H) d \lambda \wedge d\phi  +d \dot{\phi}\wedge d \phi  \right).
\label{b7a}
\een
Dualizing, we now construct the magnetic field $B_a= {1\over 2} \epsilon_{abc}\omega_{bc}$,
\ben
(B_\phi,B_{\dot{\phi}},B_\lambda)=
 e^{3 \lambda}(-\dot{\phi}/H, 3 \dot{\phi}+V_{,\phi}/H, -1),
\label{b8a}
\een
where $H =\pm\sqrt{{1\over 3} \left({1\over 2} \dot{\phi }^2 +V(\phi)\right) - k e^{-2 \lambda}}$. One can check that $B_a$ is divergence-free,
$\partial_\phi B_\phi+\partial_{\dot{\phi}} B_{\dot{\phi}}+\partial_\lambda B_\lambda=0.$
As explained at the end of the last section, the magnetic field is naturally expressed as the curl of a vector potential $A=p_i dq^i$.  Explicitly, we have
\ben
(A_\phi,A_{\dot{\phi}},A_\lambda)=
e^{3 \lambda}(\dot{\phi}, 0,-6 H),
\label{b8aa}
\een
with $H$ expressed in terms of $\phi, \dot{\phi}$ and $\lambda$ as above. 

For the purposes of our discussion, however, a more illuminating choice is to eliminate $p_\phi$. This choice is nice because now two of the remaining three variables $a$ and $H$ have monotonic properties for $k\leq 0$ making it simple to construct a surface $S$ which every classical trajectory crosses only once. From the Hamiltonian constraint ${\cal H}=0$ we obtain
\ben
p_\phi=\pm \sqrt{{1\over 6} p_a^2 a^2 -2 a^6 V(\phi)+6 a^4 k}.
\label{b5c}
\een
The square root again causes no problems: it just means the scalar field can generically have a positive, or negative velocity at each point $(\phi,H, \lambda)$ in the classically allowed domain. In fact, for this choice of variables and for simple potentials which rise monotonically from a single minimum (like $\phi^2$ or $\phi^4$), we find from (\ref{b5}) that for each $H$ and $\lambda$ the range of $\phi$ is  bounded by 
the two roots of the equation
\ben
V(\phi)= 3 (H^2+k e^{-2 \lambda}).
\label{b9}
\een

In the coordinates $\phi, H$ and $\lambda$, the velocity of the flow is:
\ben
V^a=(\dot{\phi}, \dot{H}, \dot{\lambda})=\left(\pm \sqrt{6 H^2 - 2 V + 6  k/a^2}, V - 3 H^2 - 2 k/a^2, H\right),
\label{b5d}
\een
and the canonical two-form (\ref{b6a}) is 
\bea
\omega_{C}=
e^{3 \lambda}( -6  d H \wedge d\lambda\pm 3{ 
6 H^2 -2 V +4 k e^{-2 \lambda} \over \sqrt{
6 H^2 -2 V +6 k e^{-2 \lambda}}}  d\lambda \wedge d\phi\cr
\pm
{6 H \over \sqrt{6 H^2 -2 V +6 k e^{-2 \lambda}}}
dH \wedge d \phi).
\label{b7}
\eea
Dualizing, the magnetic field is
\bea
(B_\phi,B_H,B_\lambda)= 
6 e^{3 \lambda}(-1, \pm {1\over 2}{ 
6 H^2 -2 V +4 k e^{-2 \lambda} \over \sqrt{
6 H^2 -2 V +6 k e^{-2 \lambda}}},
{\mp H\over  \sqrt{
6 H^2 -2 V +6 k e^{-2 \lambda}}}),
\label{b8}
\eea
which is parallel to the flow (\ref{b5}) and, again, divergence-free. The natural vector potential $A=p_i dq^i$ is found to be 
\bea
(A_\phi,A_H,A_\lambda)= 
e^{3 \lambda}\left(\pm
\sqrt{
6 H^2 -2 V +6 k e^{-2 \lambda}},0,-6H\right).
\label{b8aaa}
\eea

The magnetic field we have calculated describes the flux per phase space area of all classical trajectories. All that remains is to choose a suitable surface $S$ which the trajectories each cross once. As mentioned following equation (\ref{b7}), for $k\leq 0$, the Hubble parameter $H$ is monotonically decreasing. Hence it makes sense to slice phase space on a surface $H=H_S$=constant (or some mild deformation thereof). To compute the flux through the surface we must then integrate $B^H$ over the directions lying within the surface, parameterized by $\phi$ and $\lambda$, obtaining~\cite{foot}
\ben
\int \int 3 e^{3 \lambda} 
{6 H_S^2 -2 V +4 k e^{-2 \lambda} \over 
\sqrt{6 H_S^2 -2 V +6 k e^{-2 \lambda}}
} \, d\phi d\lambda\,.
\label{flu}
\een
The integral converges for negative $\lambda$ but diverges for large positive $\lambda$: this is essentially the infinity identified by Hawking and Page. However, a key point is that in the limit of large $\lambda$, the universe becomes spatially flat. In this limit, the value of $\lambda$ is neither geometrically meaningful nor physically observable. Our proposal for dealing with this physical degeneracy of solutions is to identify all universes which are indistinguishable on the chosen $H=H_S$ slice. The natural dimensionless, geometrical measure of the curvature of space is the ratio of the space curvature term to the Hubble parameter term in the Friedmann equation, $\Omega_k = -k e^{-2 \lambda_S}/H_S^2$. Our proposal is to identify universes for which $|\Omega_k|$  is smaller than some bound $\Delta \Omega_k$. In doing so, we collapse the integral over large values of $\lambda$, effectively introducing a cutoff in $\lambda$, given by $e^{2\lambda_{max}}= 1/(\Delta \Omega_k H_S^2)$. As we now show, if $\Delta \Omega_k$ is low enough and $H_S$ small enough that the expansion of the universe is adiabatic as far as the matter fields are concerned, then as far as some predictions are concerned, the cutoff dependence disappears from the result.

We are left with an integral over a two-dimensional surface with a boundary $\lambda=$ constant. Stokes' theorem, $\int {\bf B\cdot d S} = \oint {\bf A \cdot dl}$, along with equations (\ref{b8aaa}), (\ref{b3}), (\ref{b5a}) and $a=e^\lambda$, further reduce the measure to
\ben
2 \int a^3 |\dot{\phi}| d \phi \equiv \oint p_\phi d \phi,
\label{fin}
\een 
which is nothing but the standard expression for the adiabatic invariant \cite{Landau} for a homogeneous field $\phi(t)$ evolving in a time-dependent background $a(t)$. The result (\ref{fin}) could have been anticipated at the outset by expressing $\omega = d(p_a da+p_\phi d \phi)$, and noting that the first term does not contribute when $a=$ constant. It was necessary, however, to work through the intermediate steps as we have done in order to explicitly study the integral over $\lambda$ and show there is no additional surface term. Note that the $a^3$ factor in (\ref{fin}) is essential to the derived measure being conserved. One may, in the $k=0$ case, eliminate $H$ from (\ref{b3}) using (\ref{b4}) and thereby obtain a closed second order differential equation for $\phi$. This gives an autonomous first order system in the $(\phi, \dot{\phi})$ plane. However, this is {\it not} a Hamiltonian system as can be seen from the fact that it possesses an attractive fixed point. Therefore one may not use the simple measure $d\phi d \dot{\phi}$.

Now, let us justify our claim that any dependence on $\Delta \Omega_k$ disappears from the probability distribution for certain quantities, provided $H_S$ is taken 
sufficiently low that the expansion of the universe is adiabatic as far as the matter fields are concerned. The point is that in this regime the scalar field oscillates rapidly, with frequency $m \gg H$, and its stress energy is accurately described by a perfect fluid with zero pressure. The evolution of the scale factor $a(t)$ then becomes a background function of time, independent of the phase of the scalar field oscillation. We can then define the Hamiltonian for the matter field alone,
\ben
{\cal H}_\phi = {1\over 2} {p_\phi^2\over a^3(t)} + a^3(t) V(\phi),
\label{hamphi}
\een
in which $a(t)$ is treated as a background variable. When the scalar field is oscillating about the minimum of $V(\phi)$, we can approximate the potential as ${1\over 2} m^2 \phi^2$. Using this we can calculate the adiabatic invariant (\ref{fin}), obtaining
\ben
\oint p_\phi d\phi = 2 \pi m^{-1} {\cal H}_\phi \equiv 2 \pi J.
\label{adib}
\een
The ``angle" variable canonically conjugate to $J$ is then found to be
\ben
\theta= {\rm tan}^{-1} \left({ p_\phi \over m q_\phi a^3(t)}\right),
\label{angle}
\een
which obeys the following equation of motion:
\ben
\dot{\theta}= -m - {3\over 2} {\dot{a} \over a} {\rm sin} 2 \theta.
\label{angleeq}
\een

\begin{figure}[t!]
{\centering
\resizebox*{4in}{3in}{\includegraphics{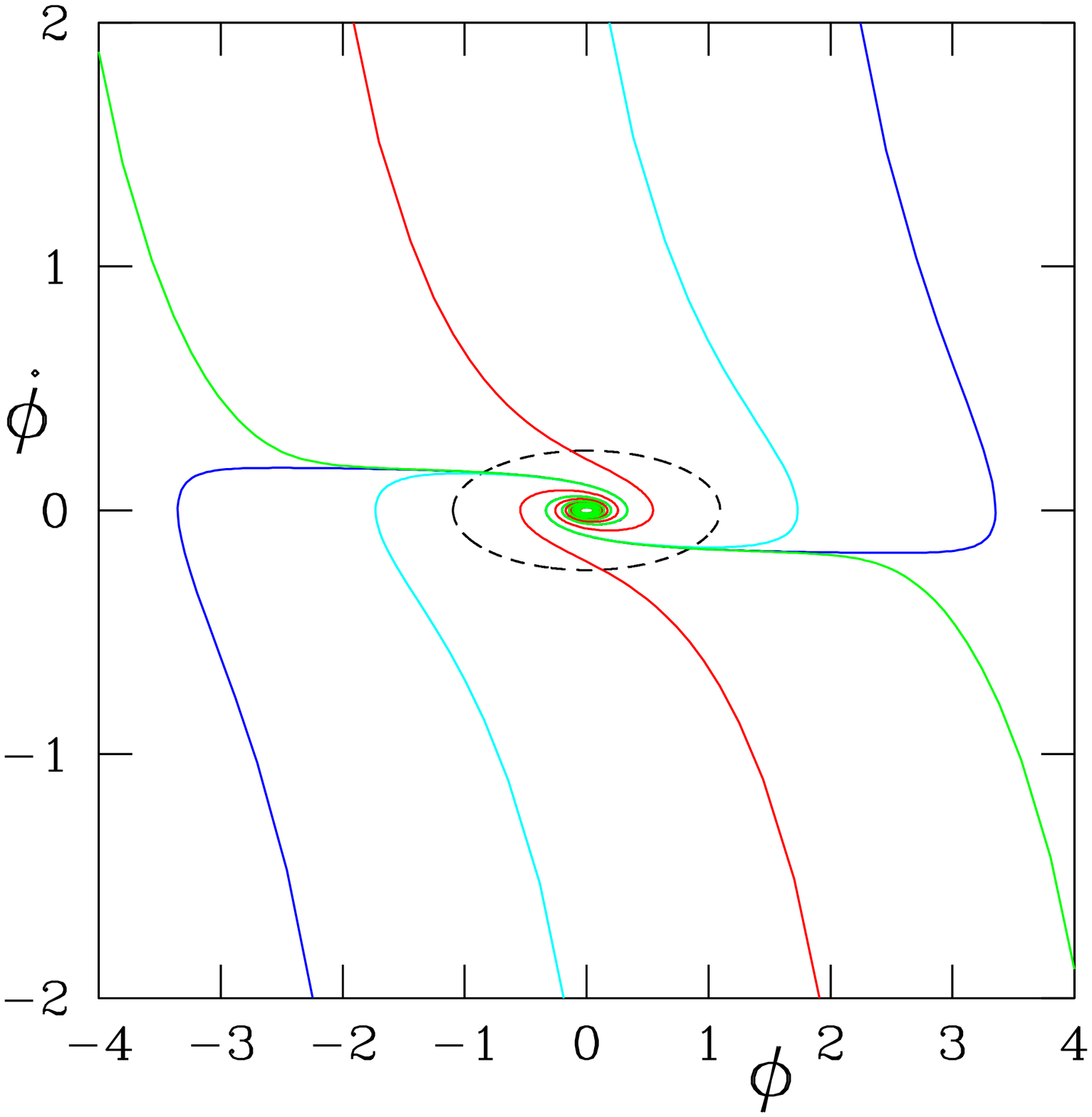}}
\resizebox*{4in}{3in}{\includegraphics{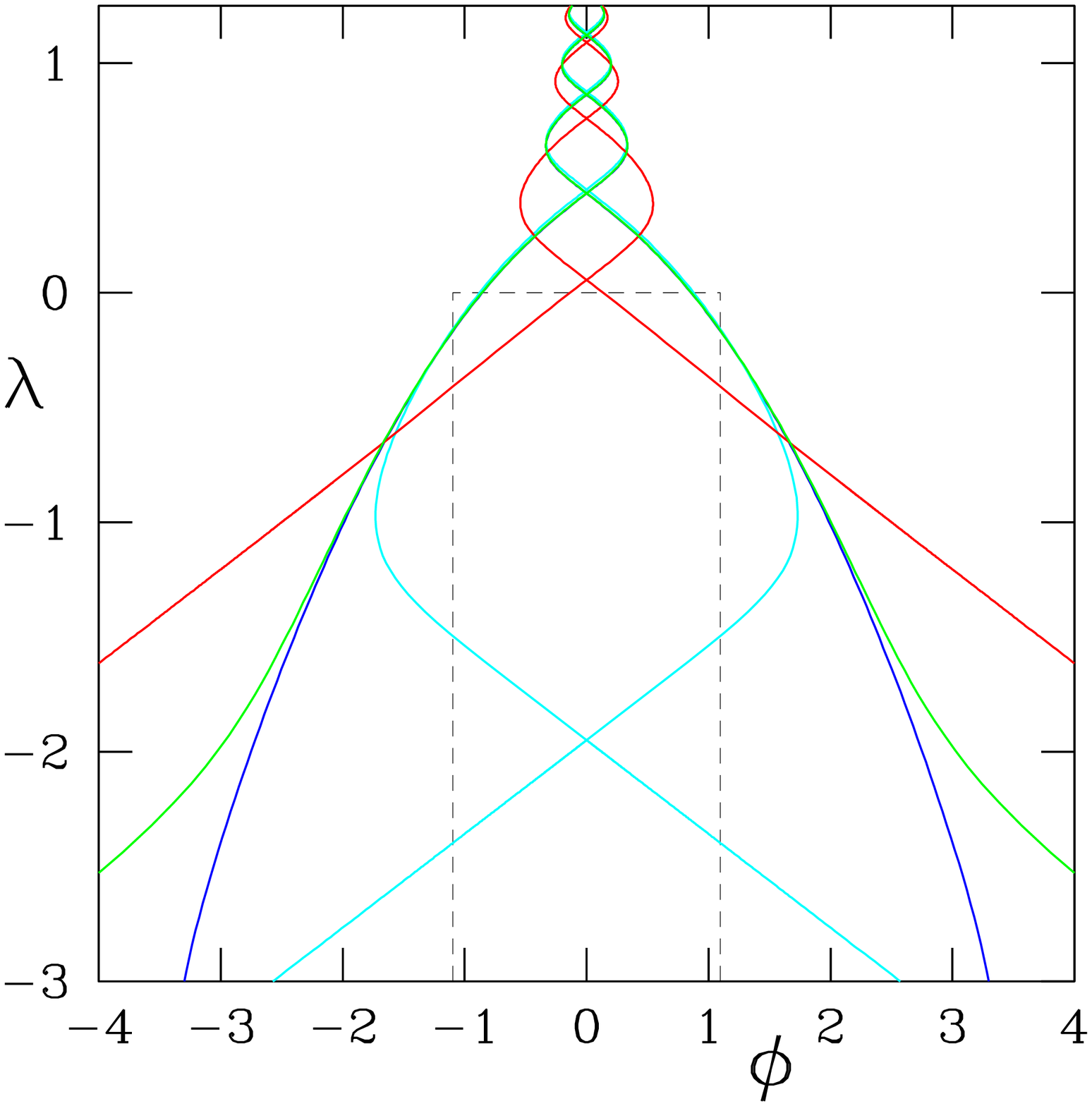}}
}
\caption{Set of classical trajectories for the inflationary model with $V={1\over 2} m^2 \phi^2$ and $k=0$, in the coordinates $(\phi,\dot{\phi}, \lambda)$, where
$\lambda \equiv \ln a$.  The upper panel shows the projection onto the $(\phi, \dot{\phi})$ plane, and the lower panel shows the projection onto the $\phi, \lambda)$ plane. The dashed lines indicate the projection of the measure surface $S$, which takes the form of an elliptical cylinder and which each trajectory crosses once. The parameters used were $m^2=0.05$, $H_S=0.1$.}
\label{ppid}
\end{figure}

In the regime of low $H$ where the the canonical measure reduces to  an adiabatic invariant, it becomes independent of $\theta$. (A similar conclusion about the behaviour of the measure at late times was reached by Starobinsky in his early work on inflation-like models~\cite{Starobinski}.) It follows that the canonical measure for $\theta$ loses its cutoff dependence in this regime. As we shall show in detail in the next section, for a universe which has undergone a substantial amount of inflation, $\theta$ must necessarily lie in an exponentially narrow range. Once inflation is over, and the expansion of the universe may be treated as adiabatic, from (\ref{angleeq}) one can show that the separation of two nearby trajectories in $\theta$ rapidly tends to a constant. Hence if we estimate the range of $\theta$ corresponding to $N$ e-folds of inflation, near the end of inflation, this estimate will remain valid for as long as the expansion (or contraction) of the universe may be treated as adiabatic. This justifies our claim that our calculated probability for $N$ e-folds of inflation becomes independent of the cutoff $\Delta \Omega_k$ in the late universe.

In contrast, the canonical measure for the scale factor $a$ (or equivalently $\Omega_k$) is strongly cutoff dependent.  To see this, note that up to a constant, $J$ is just $\rho_\phi a^3$ where $\rho_\phi$ is the effective density of matter contributed by the scalar field oscillations. Defining $\Omega_\phi= \rho_\phi/(3 H^2)=1-\Omega_k$, Friedmann's equation (\ref{b3}) yields
\ben
\Omega_\phi= 1+{k\over (a H)^2},
\label{feq}
\een
which, for fixed $H=H_S$, tends to unity for large $a$. The canonical measure,
$d (\rho_\phi a^3)$ then yields the probability distribution
\ben
\int_0^{a_c} d (\rho_\phi a^3) \propto 
\int_{a_{\rm min}}^{a_c} d\left(a^3 (1+{k\over (a H_S)^2}) \right)
\propto \int_0^{1+k\Delta \Omega_k}  
d\left( {\Omega_\phi\over |\Omega_\phi-1|^{3\over 2}}\right) 
\labeq{canmea}
\een
where $a_{\rm min}=0,H_S^{-1}$ for $k=+1,-1$ and the upper cutoff $a_c = H_S^{-1} (\Delta \Omega_k)^{-{1\over 2}}$. The resulting distribution favours flat universes, as Hawking and Page pointed out\cite{HawkingPage}, but the result is dominated by the cutoff $\Delta \Omega_k$. Note, however, that from our point of view we are not using the canonical measure as a physical theory which makes predictions, rather we are using it as a framework for assessing how predictive theories are. What we can conclude, however, is that from this point of view, a flat universe is not so surprising.

Let us illustrate these remarks with the simple inflationary model $V(\phi)= {1\over 2} m^2 \phi^2$, with $m\ll 1$ in Planck units. We start with the spatially flat case, $k=0$. Figure (\ref{ppid}) shows the evolution in the first set of variables, $\phi, \dot{\phi}, \lambda$. The upper panel is the classic plot~\cite{Belinsky,KLS} showing the attractor behavior in the  $(\phi,\dot{\phi})$ plane. The lower panel shows the projection onto $\phi, \lambda$, showing how the scale factor grows in the various solutions. As the set of trajectories rise in $\lambda$, they become more and more tightly twisted about the $\lambda$ axis. Notice in particular how all the solutions which have inflated for an extended period converge on a particular late-time oscillation phase. The dashed lines show a suitable measure surface $S$: the surface $H=H_S$= constant is an ellipse in the $\phi, \dot{\phi}$ plane, and a cylinder in the full three dimensional space. Our prescription amounts to cutting the cylinder on a surface $\lambda=$ constant, and integrating over the flux entering the cylinder below the cut. Since the magnetic field is divergence-free, this flux equals that leaving the cylinder on its upper cut surface $\lambda=$ constant. The total flux is given by a line integral of the vector potential around the ellipse. We have chosen $H_S$ to be the Hubble constant near the end of inflation, when the slow roll approximation fails. As argued above, the resulting measure on the phase of the scalar field oscillation $\theta$ becomes independent of the cutoff as $H_S$ is reduced to still lower values.

\begin{figure}[t!]
{\centering
\resizebox*{4in}{3in}{\includegraphics{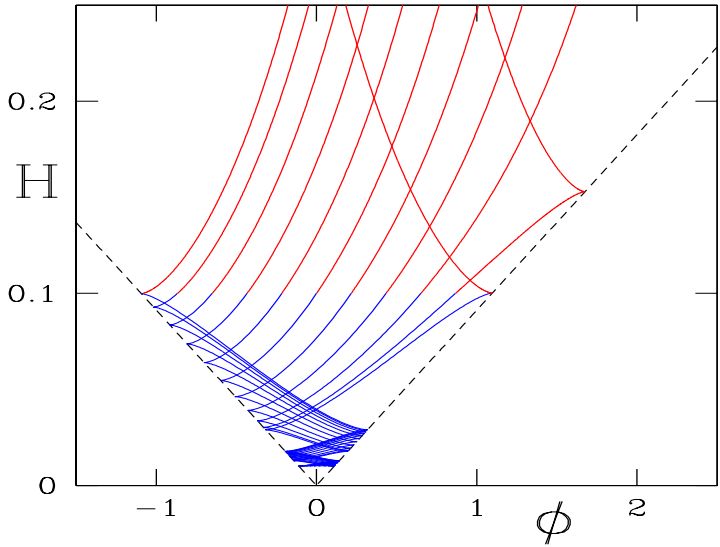}}}
\caption{The set of trajectories in the $(\phi, H)$ plane for $V(\phi)={1\over 2} m^2 \phi^2$ and $k=0$, and the same parameters as in Fig.~1. The measure surface $S$ is taken at $H=0.1$, and the trajectories plotted are equally spaced in $\phi$ on that surface. Only the trajectories with positive $\dot{\phi}$ on the measure surface are shown: those with negative $\dot{\phi}$ are obtained by mirror reflection.} 
\label{hphi1}
\end{figure}
Figure \ref{hphi1} shows the trajectories in the $(\phi,H)$ plane. As time runs forward from $S$ in an expanding universe (blue curves), all the trajectories run down in $H$ and bounce back and forth between the two boundaries $H = \pm m \phi/\sqrt{6}$ an infinite number of times.  Running time backwards from $S$ (red curves), the trajectories run to higher $H$ and generically end up kinetic-dominated. We will quantify this statement precisely in the next section.

\section{Slow-roll inflation}

We would like now to analytically estimate the measure (\ref{fin}) in models of slow-roll inflation. Slow-roll inflation requires that the derivatives of $V(\phi)$ are small: for $k=0$, the equations simplify and may be solved analytically as
an expansion in derivatives of $V(\phi)$, as follows. From equations 
(\ref{b5}) and (\ref{b7}), with $k=0$, we find
\ben
H^2= {V\over 3} + {2\over 3} ({d H\over d\phi})^2.
\label{c1}
\een
Slow-roll inflation requires the dominance of the first term on the right hand side. Provided we have a potential with a minimum and no local maximum, we can just iterate this equation to find the slow-roll solution which inflates all the way back to the Planck scale. We find this solution to be 
\bea
H_{SR}(\phi)=\sqrt{V\over 3}(1+{1\over 12}({V_{,\phi}\over V})^2+{1\over 288}(-13 ({V_{,\phi}\over V})^4+16 ({V_{,\phi}\over V})^2{V_{,\phi \phi}\over V}\hfill\cr
+{1\over 3456} (213 ({V_{,\phi}\over V})^6-432 ({V_{,\phi}\over V})^4{V_{,\phi \phi}\over V}+160 ({V_{,\phi \phi}\over V})^2 +64 {V_{,\phi}\over V}{V_{,\phi\phi\phi}\over V})+\dots)
\label{c2}
\eea
where we have implicitly assumed inflation rather than deflation. From the Friedmann equation (\ref{b5}) one sees that the sum of the sub-leading terms in this formula are due to the kinetic energy of the scalar field: one has $\dot{\phi} = 2 (d H / d\phi)$. 

The inflationary trajectory (\ref{c2}) is just one solution of the theory, and it has measure zero. What we are interested in is counting all of the solutions, and assessing how much inflation occurs in each one. This is very straightforward to do. It is convenient to take $H_S$ to be the value of the Hubble constant near the end of inflation, just before the slow-roll approximation breaks down. The measure (\ref{c2a}) then takes the simple form
\ben
\int 4 d \phi a^3 \left|{dH\over d \phi}\right|
\label{c2a}
\een
where the slope is calculated for each trajectory crossing $H_S$ and the limits of the $\phi$ integral are given by equation (\ref{b9}).

Let us now calculate the number of e-foldings of expansion which have occurred as we trace the field evolution back in time, with the scalar field rolling back up the hill to some earlier value $\phi$ which we shall take to be positive. The number of e-foldings is just $N=\int d\tau H$. From (\ref{c1}) and the Friedmann equation (\ref{b5}) with $k=0$, we find
\ben
{d N\over d \phi} = {H \over \sqrt{6 H^2 - 2 V}},
\label{c3}
\een 
which is an exact equation and does not assume the slow-roll condition.  Second, let us consider perturbing around the slow-roll solution to (\ref{c1}), given in (\ref{c2}). Setting $H\rightarrow H_{SR}+\delta H$, we find to first order
\ben
{d \delta H \over d \phi} = {3 H \delta H \over \sqrt{6 H^2 - 2 V}} =3 {dN \over d \phi} \delta H,
\label{c4}
\een
or 
\ben
{d \delta H \over d N} = 3 \delta H.
\label{c4c}
\een
This remarkably simple equation is again exact, and valid for any inflationary potential $V(\phi)$. (Similar equations were derived, for example, in Ref.~\cite{SB1}.) As we track the solution back in time, the deviation of $H$ from the slow-roll solution grows with the number of e-foldings as ${\rm exp} (3 N)$. When the deviation in $H$ becomes large, the solution departs from slow-roll. This occurs when $\delta H$ becomes of the same order as the first correction in (\ref{c2}), {\it i.e.}, when the kinetic energy becomes significantly larger than that in the slow-roll solution. If $\delta H$ is positive, as we follow the solution back in time, the kinetic energy blueshifts and quickly overwhelms the potential energy so the solution traces back to a kinetic-dominated solution with $\phi$ diverging to $+\infty$. If $\delta H$ is negative, the kinetic energy falls away to zero and the scalar field motion is turned around by the sloping potential. Again, tracking the solution back in time, the solution becomes kinetic-dominated and the scalar field diverges to $-\infty$. Slow-roll inflation occurs on the boundary between these two behaviors.

The condition for slow-roll to break down is 
\ben
\delta H = \delta H_S e^{3 N} \approx \left(H-\sqrt{V\over 3}\right)_{SR}(N) \equiv C(N),
\label{c5}
\een
where $\delta H_S$ is the deviation evaluated on our measure surface $S$, near the end of inflation, and $C(N)$ is a relatively weak function of $N$. 

For large $N$ our perturbed solution is very close to the slow-roll solution on $S$, and we can perform the integral (\ref{c2a}) to obtain the integrated probability for $N$ or more e-folds of inflation, 
\ben
P(N)\approx {\delta H_S\over {\cal N}} \approx { C(N) e^{-3 N}\over {\cal N}}, \qquad {\rm with}\quad {\cal N} \equiv \int_S d \phi |{d H \over d \phi}|.
\label{c6}
\een
This is the main result of this section. For example, if $V = m^2 \phi^2/2$, from (\ref{c2}) we find $H_{SR} \approx ({m \phi}/\sqrt{6})(1+1/(3 \phi^2)+ \dots)$ and from (\ref{c3}), $N \approx \phi^2/4$. Hence $C(N) \approx m/\sqrt{N}(1+ O(1/N))$ at large $N$. We take the measure surface to be at the value of $H$ where the slow roll approximation fails: from (\ref{c2}) we see this condition reads $\phi \approx 1$ in Planck units, which leads to $H\sim m$. The normalization factor ${\cal N}\approx m $ and we obtain $P(N)\approx N^{-{1\over 2}}{\rm exp} (-3 N)$, up to a numerical factor.

\begin{figure}[t!]
{\centering
\resizebox*{4in}{3in}{\includegraphics{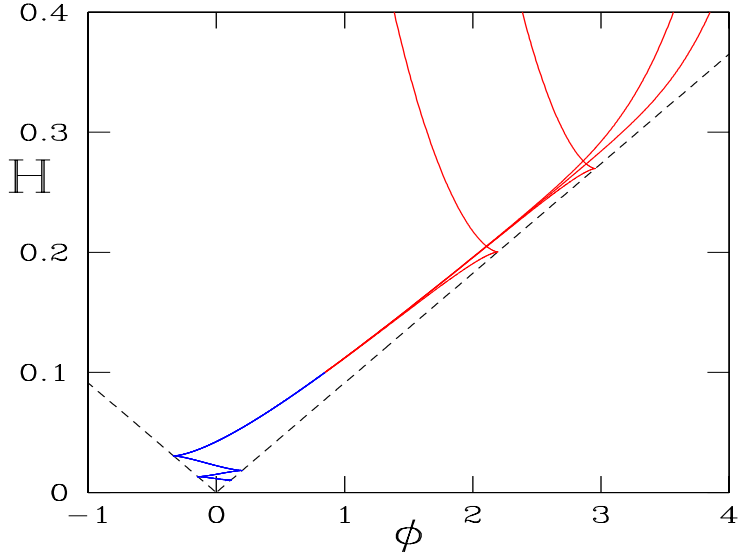}}}
\caption{Illustration of the tuning needed to create the inflationary solution as one follows $\phi$ backward in time from a measure surface $H=H_S$. Trajectories which $\phi$ is too large on the measure surface encounter the classical turning point $H=\sqrt{V(\phi)/3}$, turn around and end up kinetic dominated. Trajectories in which the initial $\phi$ is too small fall behind the inflationary solution and also head towards kinetic domination. }
\label{hphi3}
\end{figure}

Figure \ref{hphi3} shows the tuning needed to obtain a large number of inflationary e-folds. If we follow a trajectory back in time, to higher values of $H$, then the inflationary trajectory nestles close to the classical boundary $H = \sqrt{V/3}$. The corresponding initial value of $\phi$ on the surface $H=H_S$ is given by (\ref{c2}). The series converges rather slowly but a very precise value is easily obtained by a numerical ``shooting" procedure. If the initial $\phi$ is greater than the critical value, then the trajectory hits the boundary, $\phi$ reverses direction and ends up in kinetic domination, with $H \propto {\rm exp}({-\sqrt{3/2} \, \phi})$. If, on the contrary, the initial $\phi$ is smaller than the critical value, the trajectory diverges from the boundary and becomes kinetic dominated with $H \propto {\rm exp}({\sqrt{3/2} \, \phi})$.

\section{Curved Universes}

\begin{figure}[t!]
{\centering
\resizebox*{4in}{3in}{\includegraphics{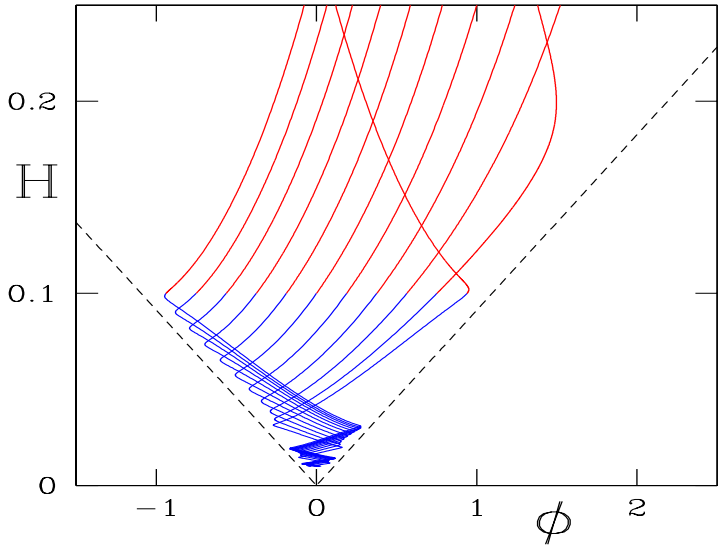}}}
\caption{Same as Figure 2, but for an open universe with $\Omega_k=0.25$ on the measure surface $H=H_S$.}
\label{hphiopen}
\end{figure}

\begin{figure}[t!]
{\centering
\resizebox*{4in}{3in}{\includegraphics{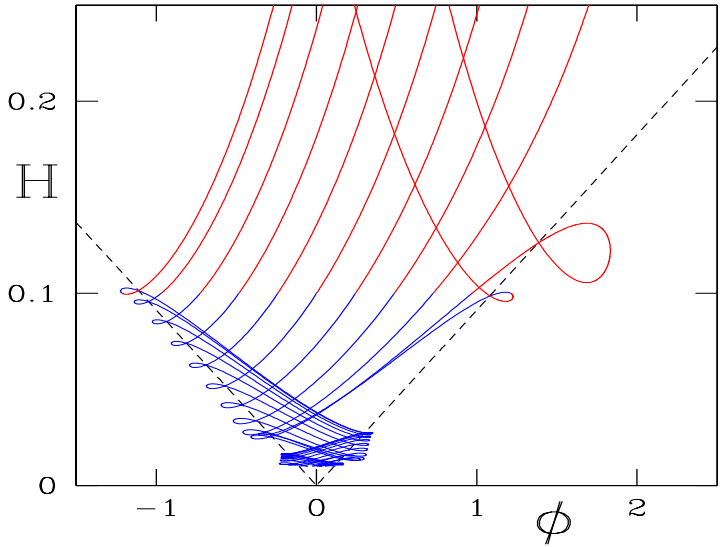}}}
\caption{Same as Figure 2, but for a closed universe with $\Omega_k=-0.25$ on the measure surface $H=H_S$.}
\label{hphiclosed}
\end{figure}

In the last section, we focused on flat universes because we found that the canonical measure possesses a divergence in the flat limit. Many of our arguments apply equally well for negatively curved universes ($k=-1$), but for positively curved universes the good monotone properties of $H$ and $a$ do not persist. Nevertheless, for the purposes of our discussion all that we really need is that the spatial curvature has a modest effect when $|\Omega_k|$ is small. This is obviously true but for the purposes of completeness, we exhibit the detailed behavior of the classical trajectories in the $(\phi,H)$ plane for nonzero  $k$. Figure (\ref{hphiopen})  shows the negatively curved case, with a modest but non-negligible value of the the spatial curvature at the measure surface, and Figure  (\ref{hphiclosed})
shows a similar picture for a positively curved universe. In both cases, it is apparent that the spatial curvature makes little difference to the kinetic-dominated trajectories when the Hubble parameter is large, since the scalar kinetic energy scales as $a^{-6}$ compared to $a^{-2}$ for the space curvature term. At lower values of $H$, the spatial curvature quickly dominates over the oscillations of the matter field,  whose density scales as $a^{-3}$. Notice the non-monotonic behavior of $H$ in the positively curved case. This has been analyzed in detail by Hawking and Page~\cite{HawkingPage}.

\section{Comparison with previous discussions}

The most recent published discussion of these issues is in the papers of Hollands and Wald~\cite{HollandsWald1,HollandsWald}, and the response by Kofman, Linde and Mukhanov~\cite{Kofman}. As we have already mentioned, because our measure is time reversal invariant, it does support Hollands and Wald's claim that deflation should be equally likely to inflation. As we have already mentioned, Hollands and Wald's discussion seems to criticize the canonical measure $\Omega_M$ on the grounds that it is hypersurface-dependent but as we have discussed in detail, this criticism does not apply to our situation, in which the surface $H=H_S=$ constant is crossed once and only once by each classical trajectory.

Kofman {\it et al.} argue that inflationary dynamics is ``definitely not" measure-preserving because energy and entropy are created during inflation. However, the relevant energy is, for us, the total Hamiltonian ${\cal H}$, which is always zero. Furthermore, the microscopic dynamics is Hamiltonian, and if one starts in a definite initial state, the fine-grained entropy is always zero. It is true that inflation is a non-adiabatic process, generating coarse-grained entropy. But the microscopic dynamics is measure-preserving and time-reversal invariant in that its definition, and indeed the definition of the multiverse $M$, does not pick out a particular direction of time. 

More specifically, Kofman {\it et al.} assume ``chaotic" initial conditions in which a closed universe is born with roughly Planckian energy density in the scalar field kinetic energy, gradient energy and potential energy. The assumption of equipartition seems to us hard to justify, since it is by no means clear that there has been time for any sort of equilibration process. Kofman {\it et al.} argue that a gradient- or kinetic-dominated closed universe would collapse in a Planck time and hence should be ignored. But, even if one insists upon considering a closed universe, as is well known, spatial inhomogeneities can allow an open universe to form within it via a process akin to bubble nucleation. Their argument does not exclude negatively curved, or nearly flat solutions which are kinetic-dominated at early times. In fact, we have found such solutions to dominate the canonical measure. Finally, 
Kofman {\it et al.} quote a formula for the probability for creation of a universe ``from nothing" which involves an {\it ad hoc}  sign flip of the usual formula for the Euclidean action. According to the usual formula, the most probable universe in fact has the {\it lowest} allowable value of $V(\phi)$, not the highest. So this argument again seems unconvincing to us.

Finally, Kofman {\it et al} discuss the ``attractor" behavior illustrated in the upper plot of Fig.~\ref{ppid}. They give an {\it ad hoc} measure which assumes Planckian initial energy density and a uniform distribution for the angle $\theta$ between the initial value of $(\phi, \dot{\phi})$ and the $\phi$ axis. They argue that all but $1-O(m)$ of the trajectories undergo inflation from the Planck era. Since $m$ is of order $ 10^{-6}$ in realistic models, they argue that inflation is virtually inevitable. However, they have omitted in their measure the $a^3$ factor in (\ref{fin}), and this means that the probability of inflation they estimate will depend very sensitively on which circle centred on the origin in $(\phi, \dot{\phi})$ space one chooses to evaluate it. Kofman {\it et al.} also consider 
imposing their measure at the end of inflation, and conclude that the probability is suppressed by $O(m)$. Hence they claim that even if one uses the late time measure, the probability of inflation is $O(10^{-6})$. This estimate does not agree with our calculations.

As we have seen, the probability of $N$ e-folds of inflation in the canonical measure is roughly ${\rm exp}(-3N)$, independent of the cutoff $\Delta \Omega_k$ or the value of $H_S$. For $60$ e-folds, a typical minimal number for realistic models, we find a probability  of approximately ${\rm exp} (-180)$.  In the $m^2 \phi^2$ model, he probability of inflation running all the way back to the Planck density is much smaller than this, approximately ${\rm exp}(-m^{-2}) \approx {\rm exp}(-10^{12})$! This may be compared to the probability one obtains from quantum cosmology \cite{HartleHawking} if one adopts the usual sign for the Euclidean action, {\it i.e.}, with probability $\propto {\rm exp} (-S_E)$ and $S_E \approx -24 \pi^2/V(\phi)$. Comparing a Planck-scale instanton, with $\phi \sim m^{-1}$, with an instanton yielding only one e-fold of inflation, with $\phi \sim 1$, one finds that the probability of Planck-scale inflation is  $\approx {\rm exp}(-m^{-2})$, which is parametrically of the same order as our classical result.

\section{Conclusions}

We have developed the canonical measure for homogeneous, isotropic universes and, we hope, clarified its geometrical structure. We applied this measure to simple inflationary models and found it to be divergent, in agreement with Hawking and Page. However, we identified the source of the divergence as being due to the geometrical and physical degeneracy of solutions which are spatially very flat, {\it i.e.}, for which $|\Omega_k| \ll 1$. Our proposal for removing the divergence is to identify universes which cannot be observationally distinguished, {\it i.e.}, for which $|\Omega_k|$ is smaller than a critical value. For this purpose, it was convenient to work in coordinates $\phi, H$ and $a$ in which the limit is easily imposed. The resulting integral over the scale factor $a$ may then be performed, leading to an integral over the matter variables alone.  We have shown that the resulting measure, when evaluated for low Hubble constant $H_S$ on the measure surface, reduces to an adiabatic invariant for the matter fields. In particular, it is independent of the conjugate ``angle" variable. It is this ``angle" variable which becomes exponentially focussed during inflation. Hence we find that the probability of inflation, evaluated with the canonical measure, is exponentially small, and this result is very insensitive to the cutoff. 

Had we instead chosen to implement our prescription at a very large value of $H_S$, one for which many e-folds of inflation are possible, we would have found, on the contrary, a strong dependence of the results on the cutoff parameter $\Delta \Omega_k$. One could choose to live with this conclusion, and indeed if $H_S$ is chosen large enough, and $\Delta \Omega_k$ small enough, one would find that most classical trajectories inflate. But there are two reasons for being suspicious about the conclusion. First, the inferred probability measure will be strongly cutoff-dependent, allowing no firm conclusions to be drawn. Second, we find it hard to justify why, if one adopts a time-dependent measure, the decision as to whether universes are or are not geometrically distinct should be made at the Planck time, rather than today. These arguments suggest that some new ingredient or dynamical principle is needed in the theory, in order to  explain why inflation began.

On the other hand, in 
an anthropic, or ``top down" approach to cosmology, the idea is to select universes on the basis of what they would be like at low $H$, not at their beginning. In that case, it seems clear that the appropriate measure should be a late-time measure like the one we have used, not an early one. The same statement would apply to any approach based on computing the probability of asymptotic ``out" states. 

Even though we have argued that the measure we have developed is reasonable, and fulfils the conditions originally proposed by Gibbons, Hawking and Stewart, we have no argument that it is unique or even physically relevant.  It seems to be a perfectly satisfactory {\it a priori} measure, that is, an unbiased estimator of our ignorance. However, its status as such is rather different from probability distributions which have arisen as a result of some equilibration process. In our case, there is no obvious physical mechanism which allows different members of the ensemble to interact, and without such interactions, the notion of equilibration is not relevant. What the canonical measure does is allow one to discuss in a quantitative way how different proposals for the big bang, or the beginning of the universe, cut down the space of classical trajectories and hence make predictions about the state of the universe today. 

The fact that the canonical measure, with what seems a sensible resolution of its divergence, strongly disfavors many e-folds of inflation, poses a serious puzzle for inflationary theory. It is important, we believe, for inflationary theory to explain why the kinetic-dominated trajectories which we have found to overwhelmingly dominate the canonical measure are somehow excluded. Indeed, the measure we have calculated is generous to inflation in that we have assumed spatial homogeneity and isotropy, and the canonical measure gives equal weight to every distinct classical solution, even those with very high potential energy density early on. The main conclusion we draw from this work is that the question of why or how inflation started remains a deep mystery, and a challenge for fundamental theory. Until that question is answered, we should remain cautious about claiming that cosmology's classic puzzles are ``solved".

%\begin{figure}[t!]
%{\centering
%\resizebox*{4in}{4in}{\includegraphics{sidepics.eps}}}
%\caption{.}
%\label{sidepics}
%\end{figure}
%
%How long inflation lasted (\ref{sidepics}).

\medskip 

{\bf Acknowledgements} We thank Stephen Hawking, Alexei Starobinsky and other participants in the first Cambridge-Mitchell Texas Conference for stimulating discussions, and Bob Wald for correspondence on this topic. This work was funded in part by PPARC (UK).


\begin{thebibliography}{99}
\bibitem{foot} Actually, with the sign conventions we have adopted, the magnetic flux through our surface is negative for expanding universes. We have adopted the sign conventions normal to classical mechanics, according to which a harmonic oscillator executes a clockwise circle in the phase plane as time runs forward. Whereas, in electromagnetism and geometry, the sign convention normally used is that the boundary of a surface is traversed in the anti-clockwise direction. In any case, the overall sign of the measure is unimportant since the non-degeneracy of $\omega$ guarantees that the sign will never change. In the following we ignore the overall minus sign.
\end{thebibliography}
\end{document}